\documentclass[runningheads]{llncs}
\usepackage{graphicx}
\usepackage{amsmath}
\usepackage{amssymb}
\usepackage{bm}
\usepackage{booktabs}
\usepackage{tabularx}
\usepackage{xcolor}
\usepackage{svg}
\usepackage{xr}
\usepackage[]{cite}
\usepackage{wrapfig}
\usepackage{textcomp}
\usepackage{hyperref}
\usepackage{chngcntr}

\begin{document}
\title{Noise as Domain Shift: Denoising Medical Images by Unpaired Image Translation}
\titlerunning{Denoising Medical Images by Unpaired Image Translation}
\author{Ilja Manakov\inst{1, 2} \and
Markus Rohm\inst{1,2} \and
Christoph Kern\inst{2} \and
Benedikt Schworm\inst{2} \and
Karsten Kortuem\inst{2} \and
Volker Tresp\inst{1, 3}}
\authorrunning{I. Manakov et al.}
\institute{Chair for Database Systems and Data Mining, LMU Munich, Germany \and 
Department of Ophthalmology, LMU Munich, Germany \and 
Siemens AG, Corporate Technology, Munich, Germany}
%
\maketitle            
\begin{abstract}
We cast the problem of image denoising as a domain translation problem between high and low noise domains.
By modifying the cycleGAN model, we are able to learn a mapping between these domains on unpaired retinal optical coherence tomography images.
In quantitative measurements and a qualitative evaluation by ophthalmologists, we show how this approach outperforms other established methods.
The results indicate that the network differentiates subtle changes in the level of noise in the image.
Further investigation of the model's feature maps reveals that it has learned to distinguish retinal layers and other distinct regions of the images.

\end{abstract}
\section{Introduction}
Medical imaging is one of the great pillars of modern diagnostics.
Clinicians rely on it to obtain information from inside the patient's body in a non-invasive way.
However, noise in the images erodes their quality and makes their interpretation difficult.
Moreover, it can cause algorithms, designed to automatically extract measurements from those images, to be inaccurate or fail outright.
In this paper, we focus on the domain of retinal optical coherence tomography (OCT)\cite{oct}, a standard diagnostic tool in ophthalmology.
Retinal OCT produces a series of 2D slices (b-scans) that display the depth profile of the retina, thus enabling clinicians to detect many sight-threatening diseases early in their progression.
The dominating type of noise in OCT is called speckle.
The speckle noise pattern depends on the imaged tissue and is highly sensitive to its position and orientation.
Since signal and speckle noise originate from the same physical process, distinguishing signal from noise is particularly challenging.
Interested readers are referred to \cite{speckle} for more details.

Current popular methods for denoising OCT scans, such as BM3D \cite{bm3d} or wavelet denoising \cite{wavelet}, neither incorporate knowledge about the OCT process nor about structures of the human eye. We argue that such knowledge should help in this task, given the complex and sample-dependent nature of speckle noise.
On the other hand, methods emerging from the field of deep learning \cite{Zhu_2017_ICCV,GAN_Goodfellow} have demonstrated precisely this ability, i.e. to learn the semantic characteristics of their input domains.
We, therefore, aim to leverage deep learning to create a method that can denoise retinal scans by utilizing knowledge it has gained about this domain.

While writing this paper, we discovered recent work from Halupka et al. \cite{Halupka} and Huang et al. \cite{Huang}, in which they investigated a different GAN-based approach to retinal OCT denoising.
Their approaches require paired training images, which can lead to problems with inaccurately registered images.
Additionally, in their works, the denoised domain is constructed by registering and averaging samples from the noisy domain.
Constructing denoised samples in this manner is not always feasible or possible and registration of images from different domains will likely not work well. 

Our approach casts denoising as a domain translation problem.
We demonstrate that, with some modifications, the cycleGAN model, introduced by Zhu et al. \cite{Zhu_2017_ICCV}, can learn a mapping between a low and high noise domain from unpaired training data.

We introduce our method, the HDcycleGAN model, in Section \ref{methods} and evaluate its performance quantitatively and qualitatively in Section \ref{quant} and Section \ref{qual}.
In Section \ref{feature_inspection}, we take a closer look at what our model has learned by inspecting its feature maps.
\section{Methodology}\label{methods}
\begin{wrapfigure}[15]{t}{0.65\textwidth}
\includegraphics[width=0.95\linewidth]{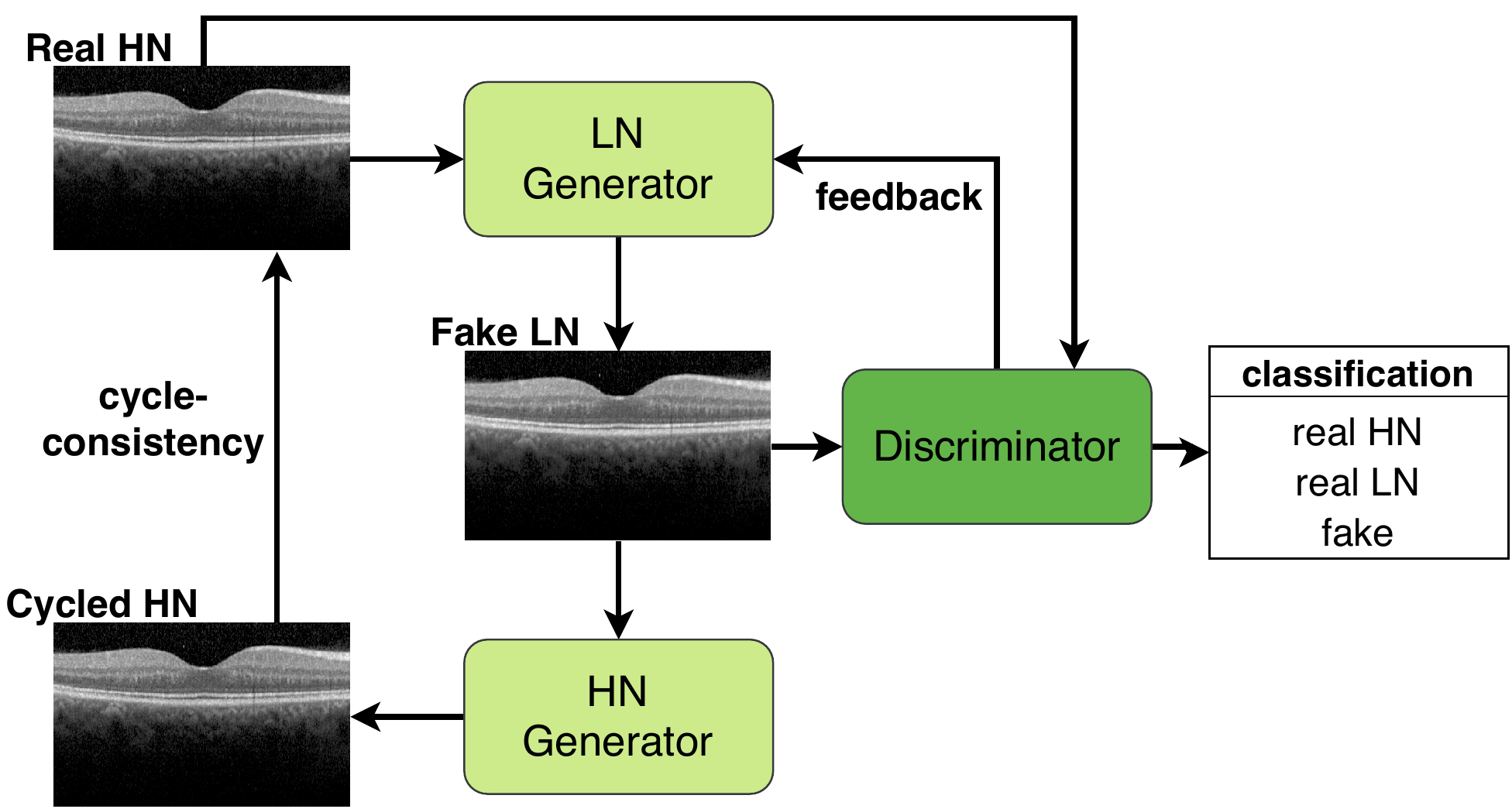}
\caption{Schematic overview of the HDcycleGAN model. The path starting from HN is shown here. Starting from LN works analogously}
\label{HDcycleGAN}
\end{wrapfigure}

Initially, we started by directly applying the cycleGAN model to the problem of learning a mapping between images of a high noise (HN) domain $\bm{h} \in H \subset \mathbb{R}^{h \times w}$ and a low noise (LN) one $\bm{l} \in L \subset \mathbb{R}^{h \times w}$.
However, we soon discovered that this model does not perform well on our problem as is.
Therefore, we made some modifications to the existing cycleGAN framework and developed our final model, the Hybrid Discriminator cycleGAN (HDcycleGAN).
Fig. \ref{HDcycleGAN} shows a pass through our model, starting from an HN image.
In the following, we briefly summarize the required knowledge about the cycleGAN and highlight the changes we made and why we made them.

The cycleGAN combines two Generative Adversarial Networks (GANs) \cite{GAN_Goodfellow} into one two-way Autoencoder.
Here, the generator of each GAN learns the mapping from one image domain to the other.
In combination, they act like encoder and decoder of an Autoencoder.
This framework allows two directions of traversal; going from domain one to domain two and back to domain one or vice versa.
The paper also introduced the cycle consistency loss, which corresponds to the reconstruction loss in the standard Autoencoder setting.
The goal of this loss function is to achieve consistency when transforming an image from one domain to the other and back.
The generators in the cycleGAN down-sample the input image using strided convolutions, pass  it through a series of residual blocks \cite{resnet} and finally use fractional-strided convolutions for up-sampling.
The discriminators down-sample their inputs through strided convolutions to produce a scalar output.

Using a cycleGAN-based approach allows us to train the network on unpaired images.
In this way, registration of images becomes obsolete and we can avoid uncertainties that arise due to interpolation in affine transformation or in cases of mismatch between the images.
An additional benefit of this framework is the cycle-consistency loss; although we are primarily interested in the mapping from HN images to LN, this added loss function provides a training signal to the network that is more stable than that of the discriminator alone.
\noindent We made three adjustments to the original cycleGAN model:
\paragraph{Skip Connections}
In our first experiments, the vanilla cycleGAN generated blurry images.
The sharpness of the image and clarity of visual features with small spatial extent play a crucial role when it comes to image quality.
To address this problem, we added skip connections to the generators, which concatenate the output of each down-sampling layer to the input of the corresponding up-sampling layer.
\paragraph{Resize-Convolutions}
Additionally, we noticed checkerboard-like artifacts in the generated images.
Following an investigation by Odena et al. \cite{odena2016deconvolution}, we replaced each fractional-strided convolution with a combination of bilinear up-sampling and a padded convolution to remedy this issue.
\paragraph{Shared Discriminator}
Even after the first two modifications and testing different hyper-parameters, the model failed to consistently improve image quality when mapping from HN to LN.
We then noticed that both discriminators learn the characteristics of real OCT b-scans independently.
The two image domains are almost identical in terms of image content.
Consequently, the discriminators could not pick up on the subtle differences between the domains (see Fig. \ref{scores} in the appendix).
As a remedy, we utilized a single discriminator that is shared between both generators. The discriminator can thus focus on the differences between the two domains instead of the full range of characteristics of each.
This change resulted in the most significant improvement in visual quality of the generated images.

This shared discriminator acts as a three-way classifier, outputting the class probabilities for real HN, real LN and fake.
As the discriminator now has to discriminate between more samples, we increased its complexity by adding a residual block with two convolutions in between each down-sampling layer.

\bigskip\noindent The loss function of our model can thus be written as follows:
Let $G_{H}: L \rightarrow H$ and $G_{L}: H \rightarrow L$ denote the generators that learn a mapping from LN to HN and from HN to LN respectively and $D: \mathbb{R}^{h \times w} \rightarrow \mathbb{R}^3$ the discriminator.
Let $\bm{t}_h, \bm{t}_l, \bm{t}_f \in \mathbb{R}^3$ be the one-hot encoded vectors that represent the classes real HN, real LN and fake. Then the loss of the network is:
\begin{equation}
\mathcal{L} = \lambda_\mathit{GAN} \left(\mathcal{L}_{G}\left(\bm{l}, \bm{h}\right) + \mathcal{L}_{D}\left(\bm{l}, \bm{h}\right)\!\right) + \lambda_\mathit{cycle} \mathcal{L}_\mathit{cycle}\left(\bm{l}, \bm{h}\right) \textrm{, with}
\end{equation}
\begin{equation}
\mathcal{L}_{G}(\bm{l}, \bm{h}) =  -\sum_{j=1}^3 \bm{t}_{hj}\log(D(G_{H}(\bm{l}))_j) -\sum_{j=1}^3 \bm{t}_{lj}\log(D(G_{L}(\bm{h}))_j)
\end{equation}
\begin{align}
\begin{split}
\mathcal{L}_{D}(\bm{l}, \bm{h}) = &- \sum_{j=1}^3 \bm{t}_{fj}\log(D(G_{H}(\bm{l})_j) - \sum_{j=1}^3 \bm{t}_{fj}\log(D(G_{L}(\bm{h})_j) \\ &- \sum_{j=1}^3 \bm{t}_{hj}\log(D(\bm{h})_j) - \sum_{j=1}^3 \bm{t}_{lj}\log(D(\bm{l})_j)
\end{split}
\end{align}
\begin{equation}
\mathcal{L}_{cycle}(\bm{l}, \bm{h}) = \Vert \bm{l} - G_{L}(G_{H}(\bm{l})) \Vert_1 + \Vert \bm{h} - G_{H}(G_{L}(\bm{h})) \Vert_1
\end{equation}
Here $\lambda_{GAN}$ and $\lambda_{cycle}$ are hyper-parameters for weighting discriminator and cycle-consistency loss respectively.
For our model, we set $\lambda_{GAN}=1$ and $\lambda_{cycle}=10$ following \cite{Zhu_2017_ICCV}. Our implementation of the described methodology is publicly available at \href{https://github.com/IljaManakov/HDcycleGAN}{github.com/IljaManakov/HDcycleGAN}.
We also provide implementation details in Tables \ref{generator} and \ref{discriminator} in the appendix.
\section{Experiments and Results}
After training the HDcycleGAN for 245 epochs with an Adam optimizer and a learning rate of $5 \times 10^{-4}$, we performed both quantitative and qualitative analyses on the test set, which we explain in sections \ref{quant} and \ref{qual}.
In the quantitative analysis, we compared our approach to popular denoising methods using several measurements of similarity between real LN images and denoised ones.
For the qualitative analysis, the similarity between real LN images and images produced by BM3D \cite{bm3d}, wavelet denoising \cite{wavelet} and our method was assessed by three ophthalmologists independently.
Finally, in section \ref{feature_inspection}, we inspect the learned feature maps of the LN generator. We start by describing our dataset.
\subsection{Dataset}
We acquired the data for this task in-house, using a SPECTRALIS OCT+HRA from Heidelberg Engineering, as part of the general diagnostic workflow for macular diseases.
We did not select patients based on any further traits.
As such, the scans in the dataset show various kinds of diseases in all stages and are representative of the typical imaging data generated at our hospital.
To gather the images belonging to the high noise domain, we followed the hospital protocol, using 30\textdegree ART Volume acquisition with 12 frames averaged for each b-scan.
The low noise domain consists of acquisitions that follow the same protocol except that we set the number of averaged frames to 60.
We obtained both HN and LN images from the same patients on the same visit.
As the proprietary software of the device manufacturer handles the frame averaging, we did not have access to the individual frames.
In total, we gathered 23030 b-scans in 470 volumes from 235 patients for each noise domain. We used 90\% of the volumes for training and the remaining 10\% for testing.
Before passing the images through our model, we scaled the 496x512 images to a pixel intensity range between 0.0 and 1.0.
\subsection{Quantitative Evaluation} \label{quant}
To asses our model's performance, we evaluated the similarity between the generated images and the ground truth LN images in the test set.
Since we acquired HN and LN scans pairwise, we registered the images employing a registration algorithm based on discrete Fourier-transform \cite{imreg-dft}.
After registration, we calculated the peak signal-to-noise ratio (PSNR) and structural similarity index (SSIM) between the two images.
Additionally, we used the Marching Cubes algorithm \cite{marching_cubes} to find the contour of the retina.
Inverting the selection yields a background mask, while reapplying Marching Cubes on the retinal layers with a different level finds contours in highly reflective parts of the retina.
We designated these regions as signal.
This process is illustrated in Fig. \ref{sig_bg} in the appendix.
Using the signal and background regions, we then calculated the mean-to-standard-deviation ratio (MSR) and contrast-to-noise ratio (CNR).
To better gauge the performance of our approach, we included median filtering, wavelet denoising \cite{wavelet}, bilateral filtering \cite{bilateral}, non-local means \cite{nl-means} and BM3D \cite{bm3d} in the comparison.
The results are displayed in Table \ref{quant_results}.
\newcolumntype{Y}{>{\centering\arraybackslash}X}
\begin{table}
\begin{center}
\caption{Results of the quantitative analysis. Values are shown as mean $\pm$ standard deviation.}
\label{quant_results}
\begin{tabularx}{\textwidth}{l Y|Y|Y|Y}
\toprule
method	&	CNR	&  MSR	&  PSNR	&  SSIM\\
\midrule
raw 	& 	3.66 $\pm$     2.21 &      3.96 $\pm$     1.73 &      21.99 $\pm$      1.33 &       0.662 $\pm$      0.055 \\
median 		&	3.82 $\pm$     2.36 &      4.25 $\pm$     1.92 &      22.32 $\pm$      1.45 &       0.682 $\pm$      0.051\\
wavelet \cite{wavelet}  	&	3.81 $\pm$     2.37 &      4.23 $\pm$     1.86 &      22.34 $\pm$      1.41 &       0.690 $\pm$      0.053\\
bilateral \cite{bilateral} 	&	3.78 $\pm$     2.33 &      4.28 $\pm$     1.93 &      22.29 $\pm$      1.40 &       0.690 $\pm$      0.053\\
nl-means \cite{nl-means} 	&	3.78 $\pm$     2.33 &      4.43 $\pm$     2.12 &      22.32 $\pm$      1.40 &       0.702 $\pm$      0.051\\
BM3D \cite{bm3d} 		&	3.87 $\pm$     2.44 &      4.39 $\pm$     1.97 &      22.50 $\pm$      1.45 &       \bf{0.708 $\pm$      0.052}\\
ours		&	\bf{4.00 $\pm$     2.51} &      \bf{4.73 $\pm$     2.23} &      \bf{22.58 $\pm$      1.41} &       0.706 $\pm$      0.050\\
\bottomrule
\end{tabularx}
\end{center}
\end{table}
We can see that our model outperforms the other methods in all measurements except SSIM, where BM3D is slightly ahead.
Overall we find that the performance of BM3D and our model is very close in inter-image measurements (SSIM and PSNR).
In intra-image measurements (CNR and MSR) the margin between our approach and the others widens.
It is also worth noting that our algorithm requires 30\% less time to run on CPU than BM3D and beats all other algorithms by almost an order of magnitude on a low-end GPU (see Fig. \ref{run_times} in the appendix).
Although PSNR, SSIM, MSR and CNR are standard metrics of image quality, there is a caveat to these results; since HN and LN samples stem from independent acquisitions the noise in them is uncorrelated.
This might explain why the overall improvement in these metrics is relatively low for all algorithms. 
\subsection{Qualitative Evaluation} \label{qual}
\begin{figure}
	\centering
	\includegraphics[width=0.68\textwidth]{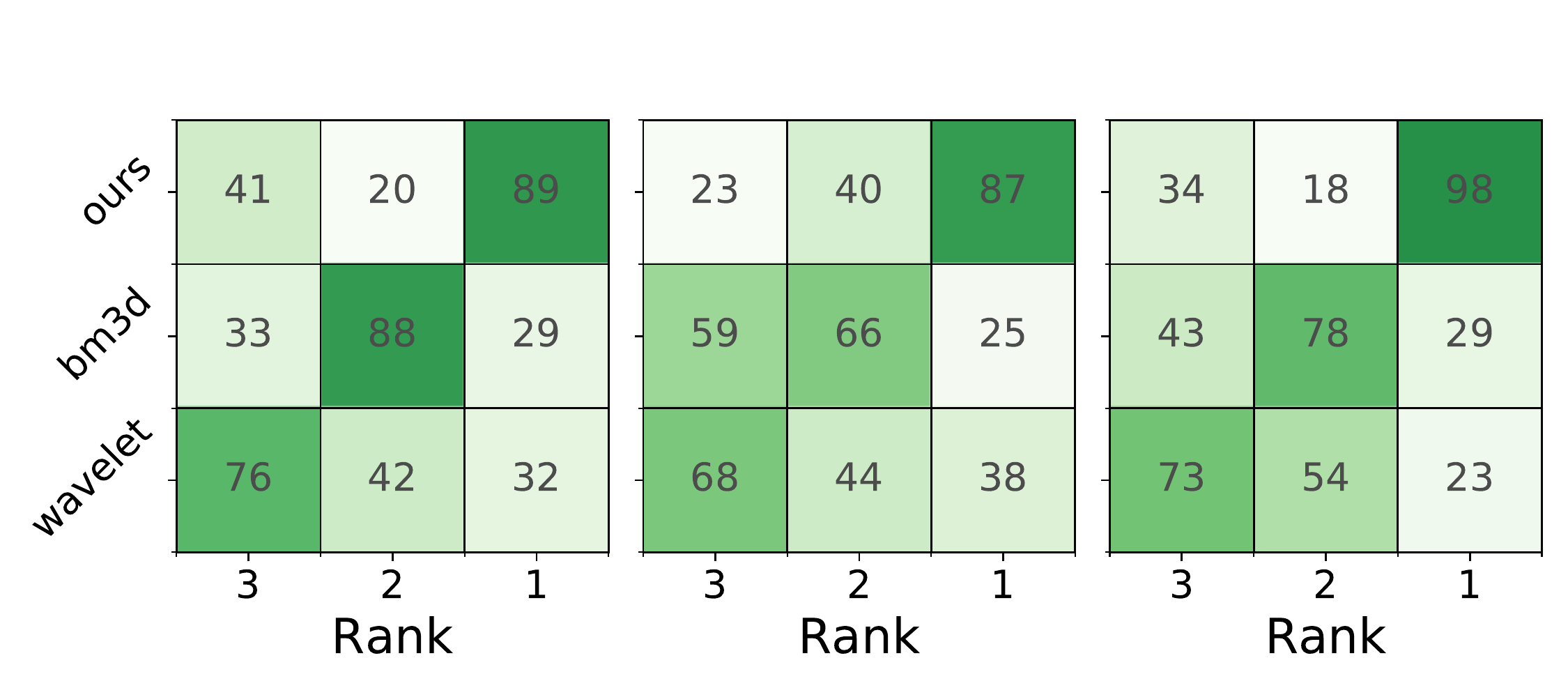}
	\caption{Results of the qualitative evaluation by three experts.}
	\label{qual_results}
\end{figure}

Because of this caveat, we asked three expert ophthalmologists to visually assess the quality of our results.
We provided them with 150 real LN images from the test set and images generated from the corresponding real HN images using BM3D, wavelet denoising and our method.
For each such sample, the clinicians rated the methods by their similarity to the real LN images.
We ordered the images in each sample randomly and did not provide any indication as to which model generated which image.
The results of this evaluation, displayed in Fig. \ref{qual_results}, confirm the findings of the quantitative evaluation.
The experts unanimously agree that our approach outperforms the benchmarks.
\subsection{Feature Map Inspection} \label{feature_inspection}
We attempted to understand how the model is approaching the task of image enhancement by looking at the feature maps that it has learned.
We did this by passing a sample through the LN generator and extracting the neuron activations at every layer.
Due to the convolutional nature of the generator, these layer outputs are shaped like images with many channels.
Hence we can view each channel in the activations of a layer as a gray-scale image which we refer to as a feature map.
\begin{wrapfigure}[20]{tr}{0.46\textwidth}
	\includegraphics[width=0.95\linewidth]{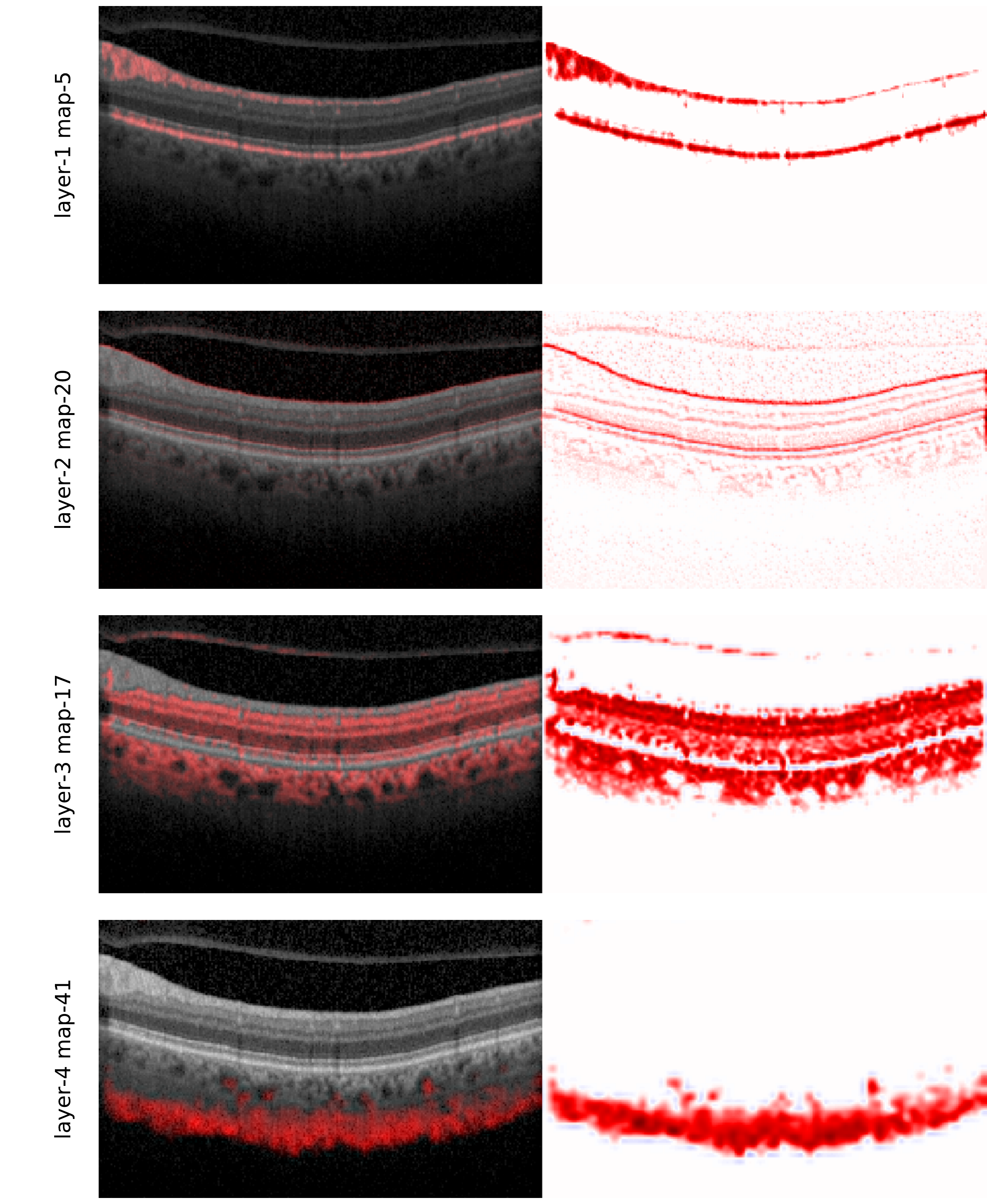}
	\caption{Example of the feature maps. On the left, the input image is overlaid with the map. On the right, the feature map is shown by itself.}
	\label{feature_maps}
\end{wrapfigure}
%
By up-scaling the feature maps to the size of the input, we then checked for spatial correlations.
For visualization purposes, we show some feature maps from different layers, which highlight distinct regions of the retina, in Fig. \ref{feature_maps}.
Many more can be found in the appendix.

We observed that the feature maps at the output of the deeper residual blocks become increasingly abstract and spatially uncorrelated with the input.
The feature maps at the outputs of the first four layers (initial convolution and down-sampling 1 to 3) and shallower residual blocks exhibit a strong spatial correlation with the input.
Moreover, the different channels seem to correspond to anatomically distinct regions in the b-scan, although segmentation was never part of the training objective.

We think that this finding is relevant when viewed from two perspectives.
Firstly, it shows that the model has gained some domain specific knowledge about the structure of macular OCT scans, which general methods such as BM3D and wavelet denoising are lacking.
Secondly, this property can prove useful from the viewpoint of transfer learning, i.e. when applying this model to other tasks.
The feature maps themselves can also be used for other tasks.

An example of the second point can be found in Fig. \ref{skeletons} in the appendix.
We discovered a feature map that appears to track the positions of the Inner Limiting Membrane (ILM) and the Retinal Pigment Epithelium (RPE) (the inner- and outermost layers of the retina) (see Fig. \ref{map114}).
We then multiplied the feature map with its corresponding b-scan, applied the image mean as a threshold and skeletonized the remainder.
The resulting lines can be used to estimate retinal thickness.
This method seems to work well even in the presence of pathologies, such as myopia (row 4, col. 4) or vitreous detachment (row 1, col. 4 and row 5, col. 3), which are typical causes for segmentation errors in commercial segmentation algorithms.
\section{Discussion}
In this paper, we applied the HDcycleGAN model to the problem of image enhancement.
In medical imaging, reduced image noise typically comes at the cost of increased acquisition time, radiation dose or other detrimental effects.
Our model can learn a mapping between domains that correspond to different settings of those costly acquisition parameters.

Additionally, our approach learns the structural characteristics of the medical imaging domain, which further improves its usefulness as it can be leveraged for other tasks in that domain.
As part of future work, we wish to study the transferability of our approach to other imaging modalities, such as Ultrasound.

As is the case with all GAN-based methods, the training of this model is not straightforward and the performance does not appear to increase monotonically throughout training.
Nevertheless, our approach allows us to pre-train the parts individually; the generators as Autoencoders and the discriminator as a classifier between domains.
In the future, we also plan to test if pre-training can improve training stability and model performance.
\bibliography{bibliography_short}
\bibliographystyle{splncs04}

\newpage
\appendix
\counterwithin{figure}{section}
\counterwithin{table}{section}
\section{Appendix}
\begin{figure}
	\label{scores}
	\includegraphics[width=\textwidth]{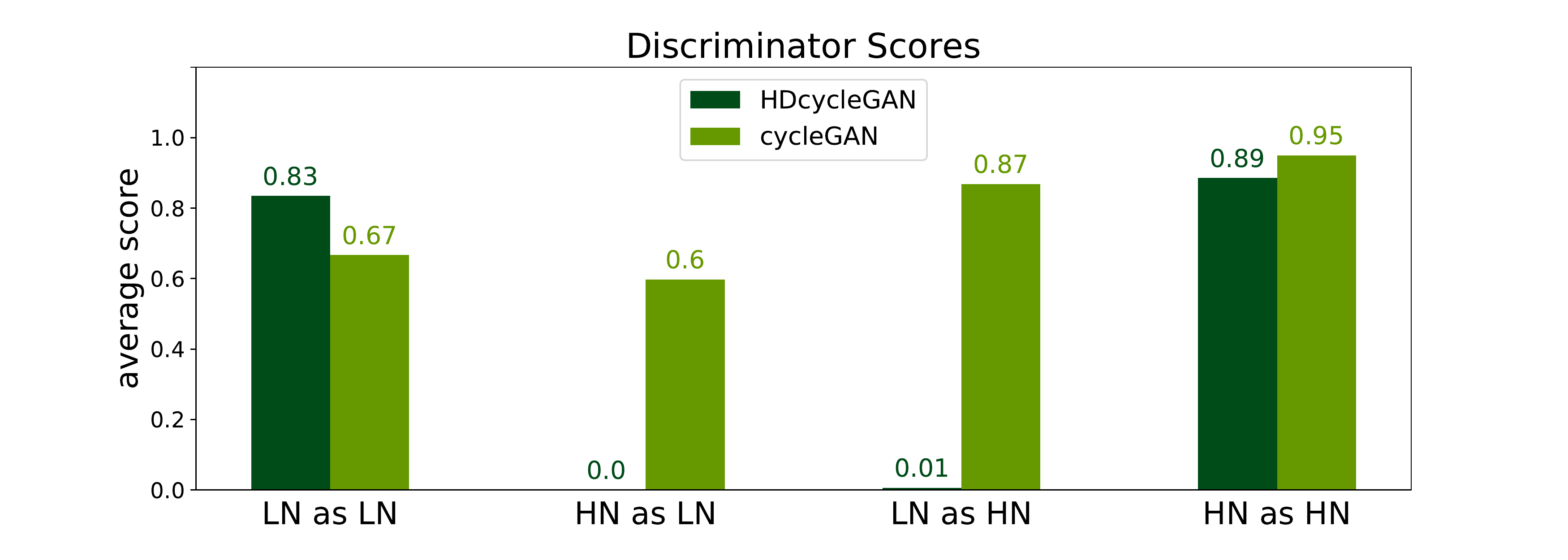}
	\caption{Mean scores assigned by the discriminators to data samples. The bars correspond to samples being evaluated for a specific target domain. In the case of HDcycleGAN the values are obtained from the class probabilities, while for the cycleGAN they are taken from the respective discriminators. We can see that in the cycleGAN the scores do not change much when we evaluate a real image as either LN or HN.}
\end{figure}

\begin{figure}
\includegraphics[width=\textwidth]{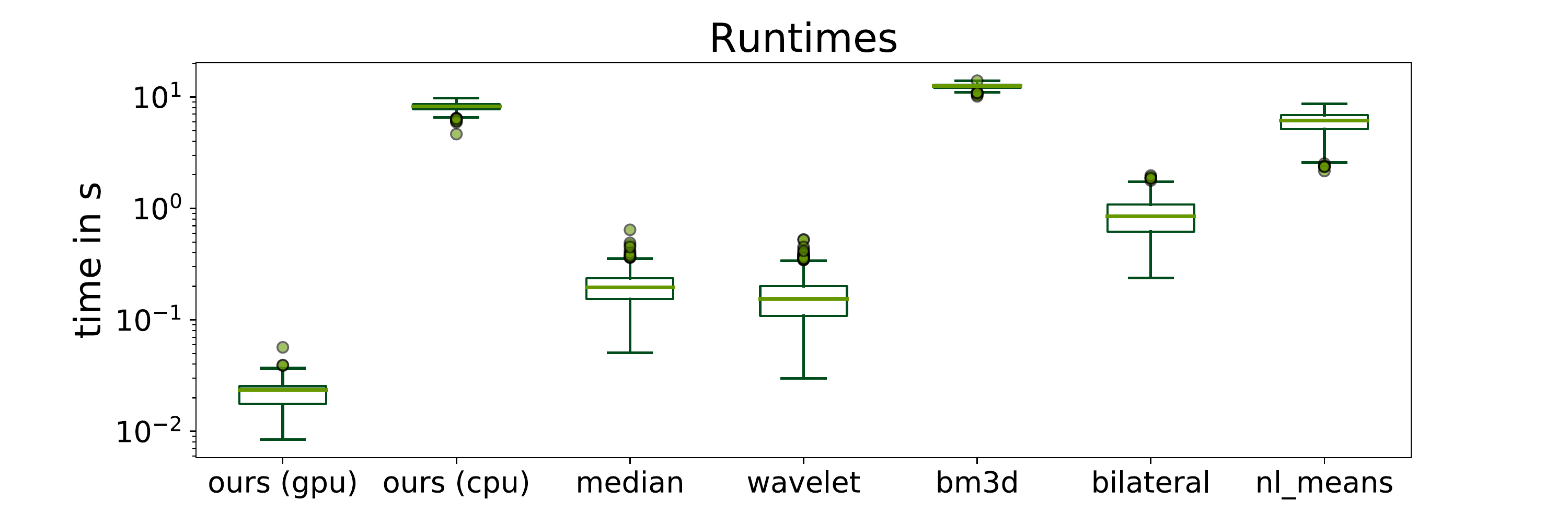}
\caption{Runtimes of the different methods computed over the test set. The GPU used for this calculation was a mobile 2GB GTX 1050.}
\label{run_times}
\end{figure}

\begin{figure}
\includegraphics[width=\textwidth]{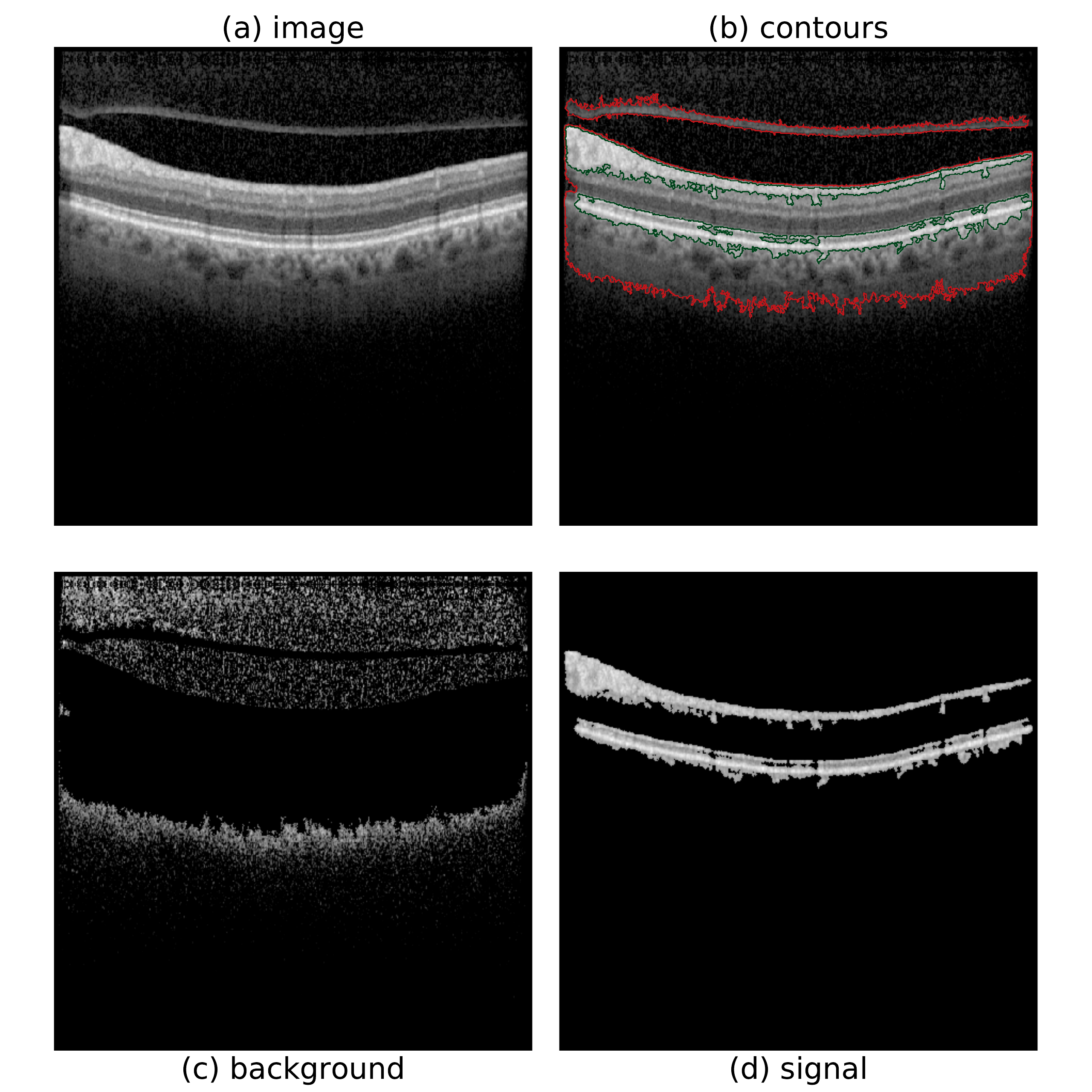}
\caption{Illustration of the process to obtain background and signal masks for calculation of CNR and MSR.}
\label{sig_bg}
\end{figure}

\begin{figure}
\centering
	\includegraphics[width=\textwidth]{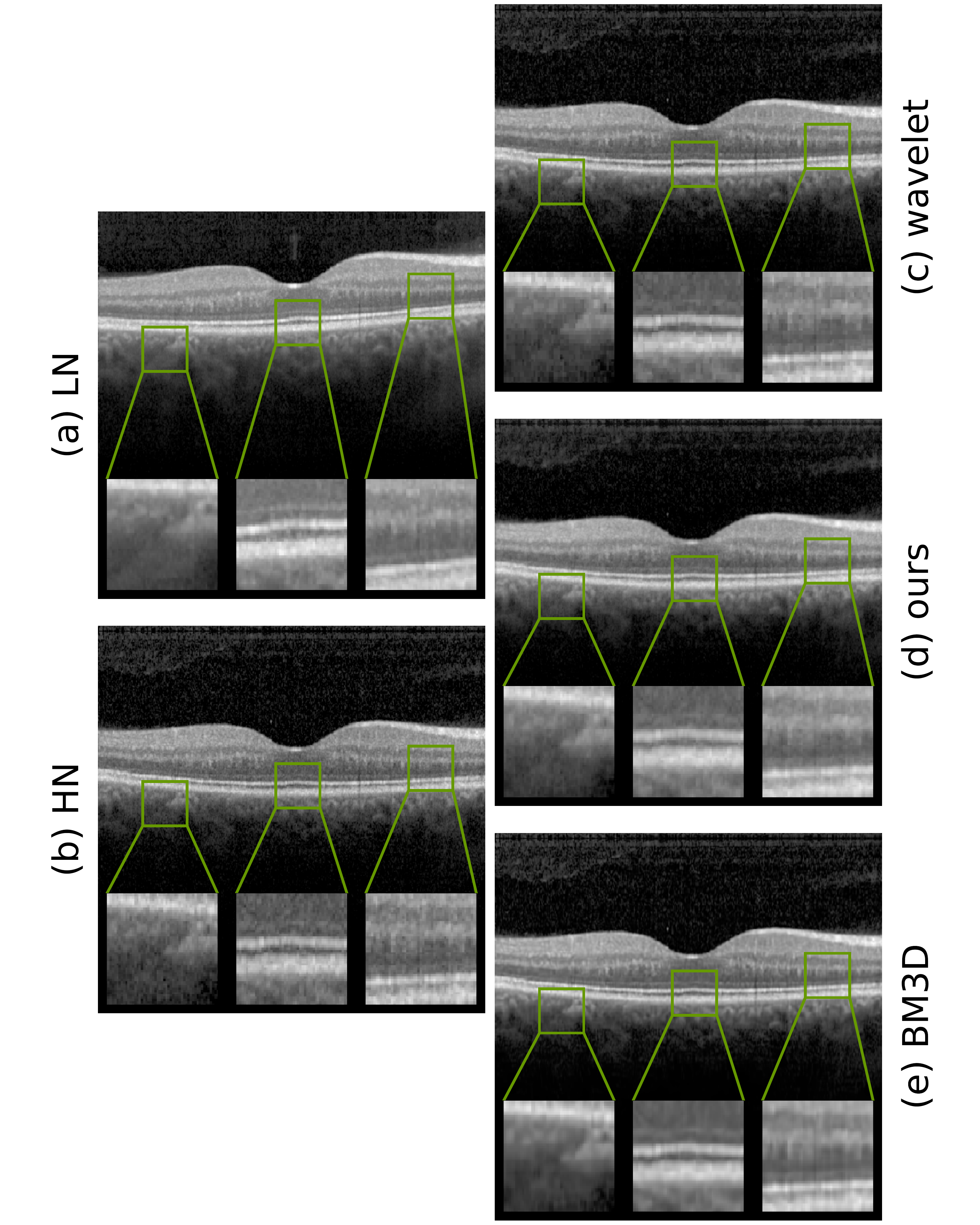}
	\label{sample}
	\caption{Sample showing a b-scan with 60 frames (a) and 12 frames (b) averaged. Also shown are the results of denoising using wavelet (c), HDcycleGAN (d) and BM3D (e)}
\end{figure}

\begin{figure}
\centering
	\includegraphics[width=\textwidth]{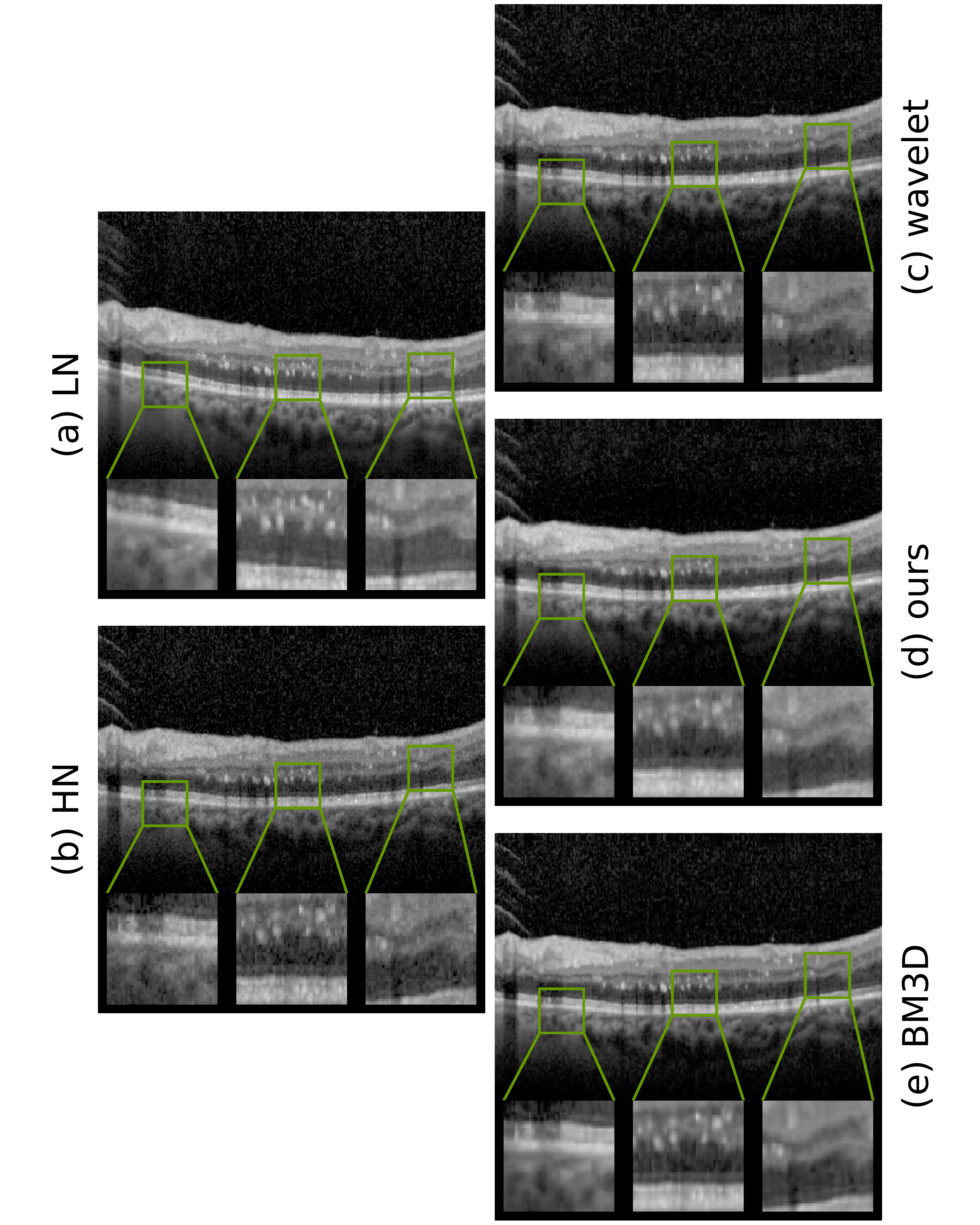}
	\label{sample2}
	\caption{Sample showing a b-scan with 60 frames (a) and 12 frames (b) averaged. Also shown are the results of denoising using wavelet (c), HDcycleGAN (d) and BM3D (e)}
\end{figure}

\begin{figure}
\centering
	\makebox[0pt]{
	\includegraphics[width=1.3\textwidth]{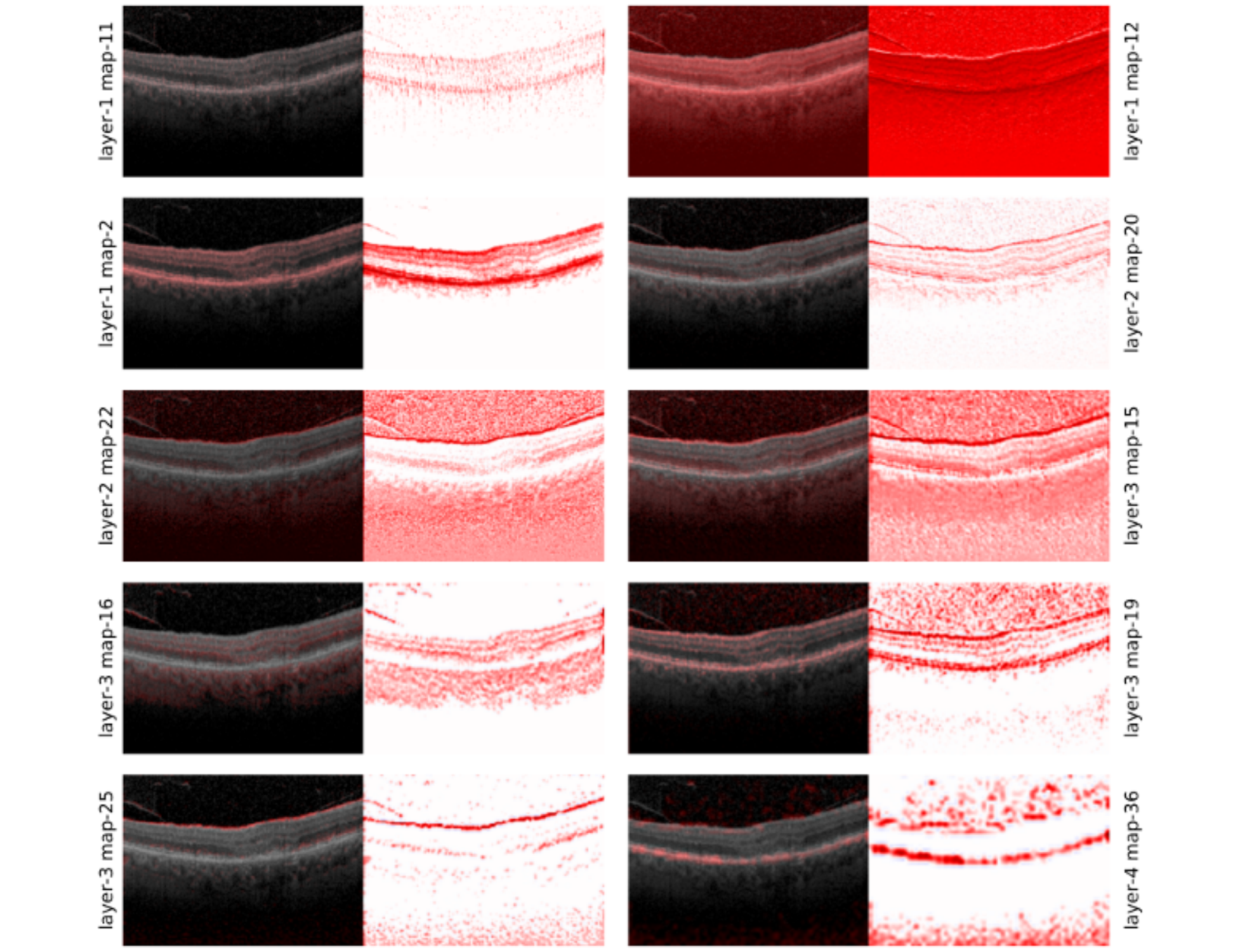}}
	\label{subj36}
	\caption{Demonstration of feature maps, part 1. The feature maps represent individual channels of activations in the LN generator at different layers. On the left, the input image is overlayed with the map. On the right, the feature map is shown by itself. Layer and channel are shown on the sides.}
\end{figure}

\begin{figure}
\centering
	\makebox[0pt]{
	\includegraphics[width=1.3\textwidth]{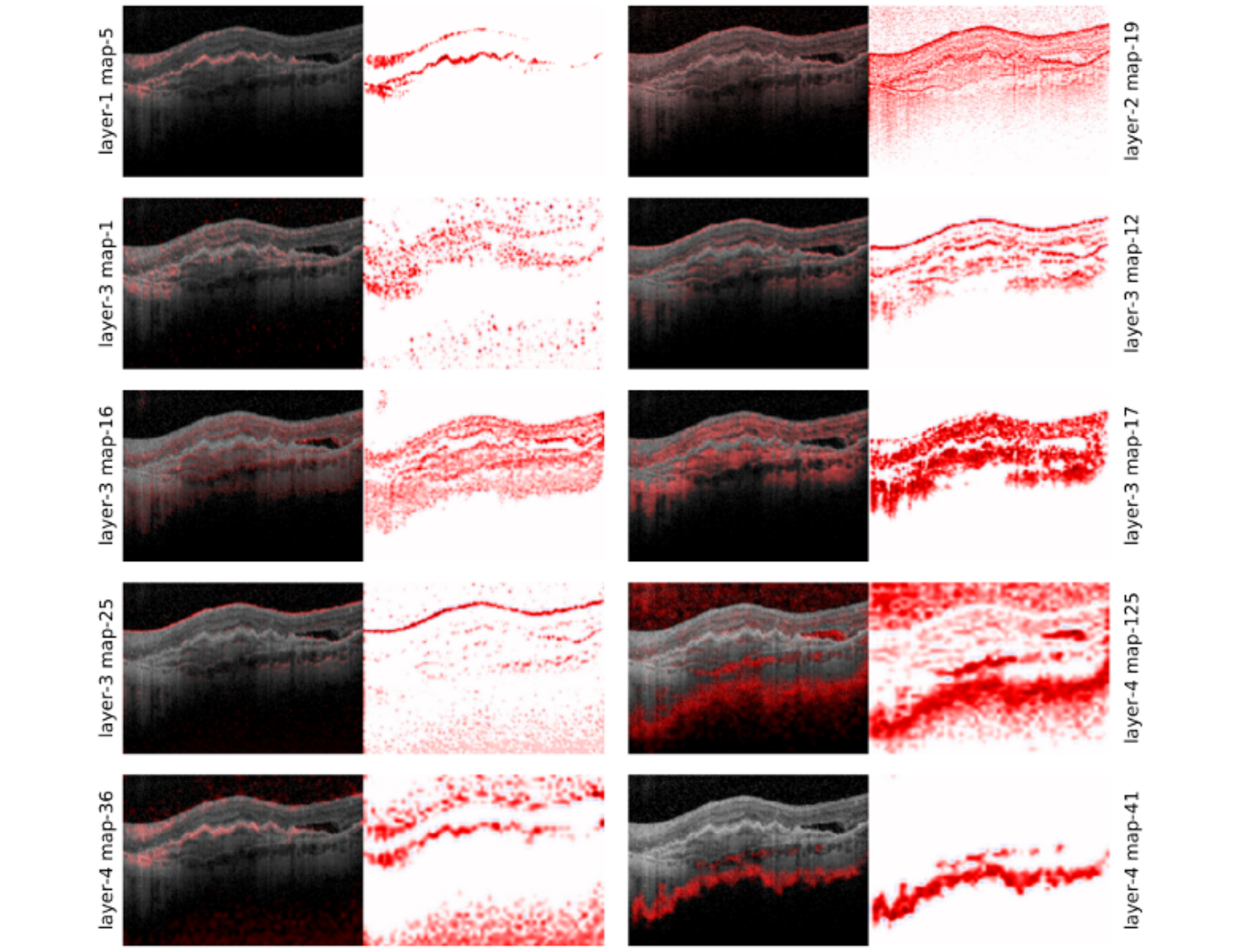}}
	\label{subj648}
	\caption{Demonstration of feature maps, part 2. The feature maps represent individual channels of activations in the LN generator at different layers. On the left, the input image is overlayed with the map. On the right, the feature map is shown by itself. Layer and channel are shown on the sides.}
\end{figure}

\begin{figure}
\centering
	\makebox[0pt]{
	\includegraphics[width=1.3\textwidth]{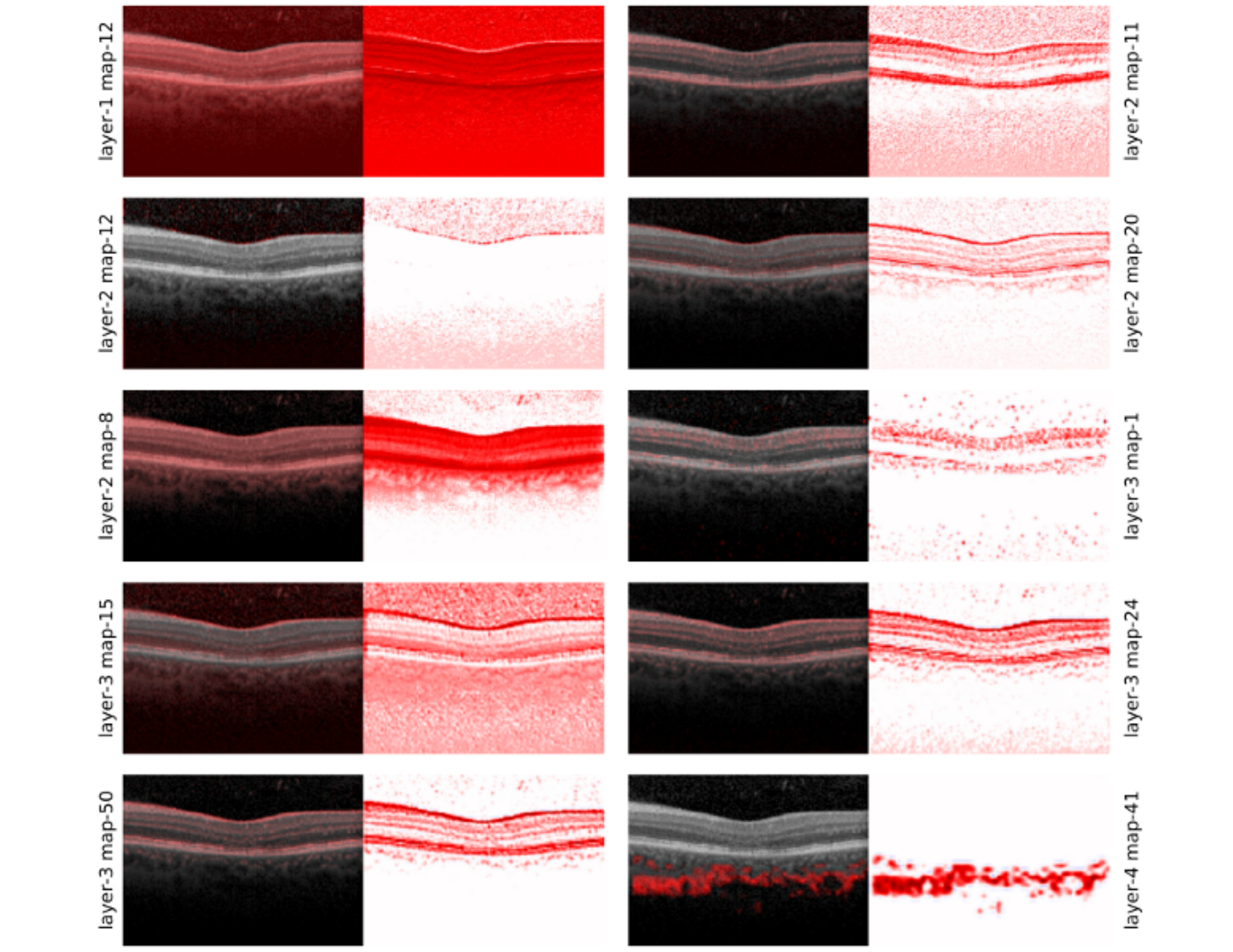}}
	\label{subj720}
	\caption{Demonstration of feature maps, part 3. The feature maps represent individual channels of activations in the LN generator at different layers. On the left, the input image is overlayed with the map. On the right, the feature map is shown by itself. Layer and channel are shown on the sides.}
\end{figure}

\begin{figure}
\centering
	\makebox[0pt]{
	\includegraphics[width=\textwidth]{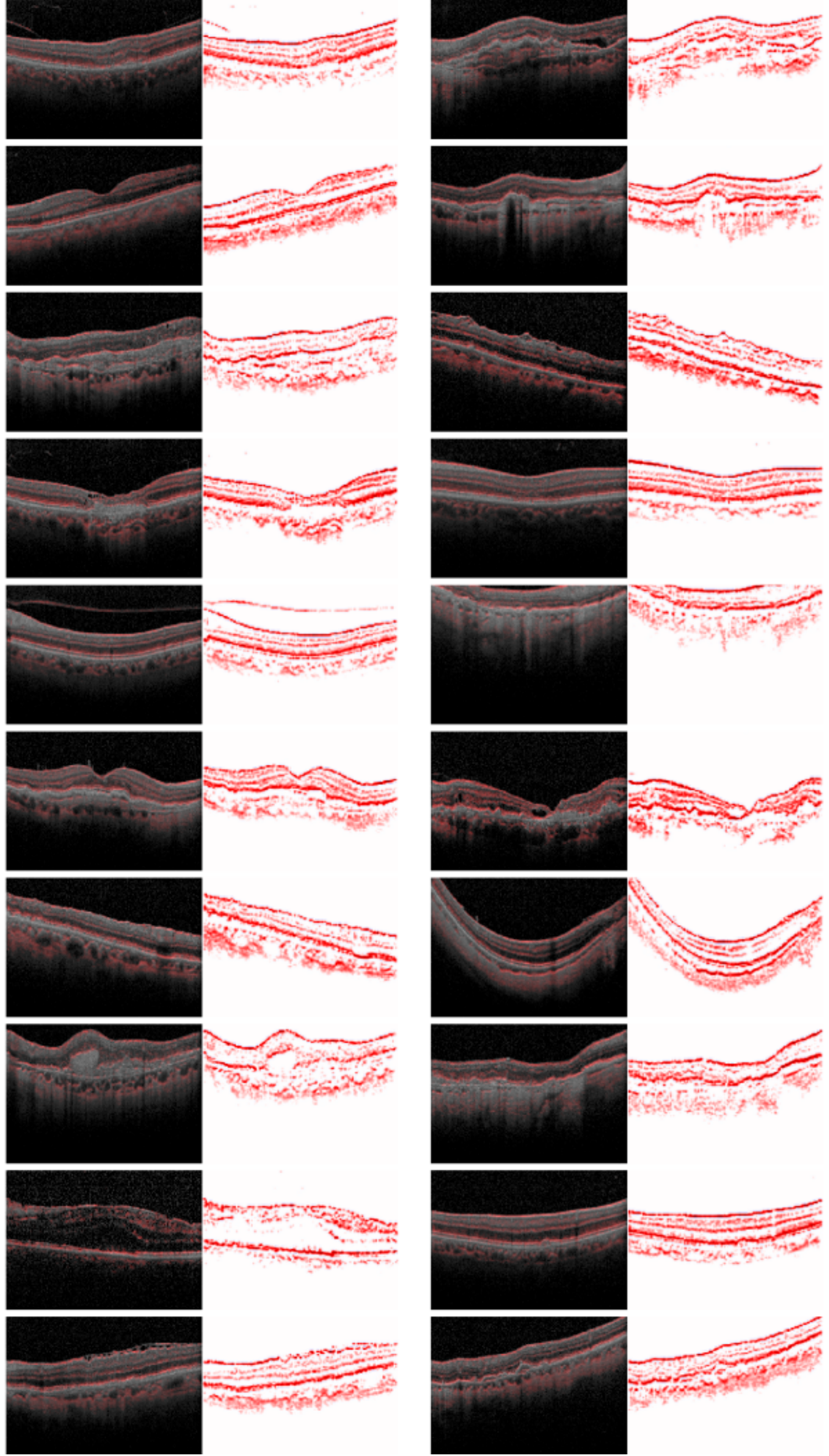}}
	\label{map12}
	\caption{Feature map 12 of layer 3 appears to have learned to detect the borders of the retinal layers.}
\end{figure}

\begin{figure}
\centering
	\makebox[0pt]{
	\includegraphics[width=\textwidth]{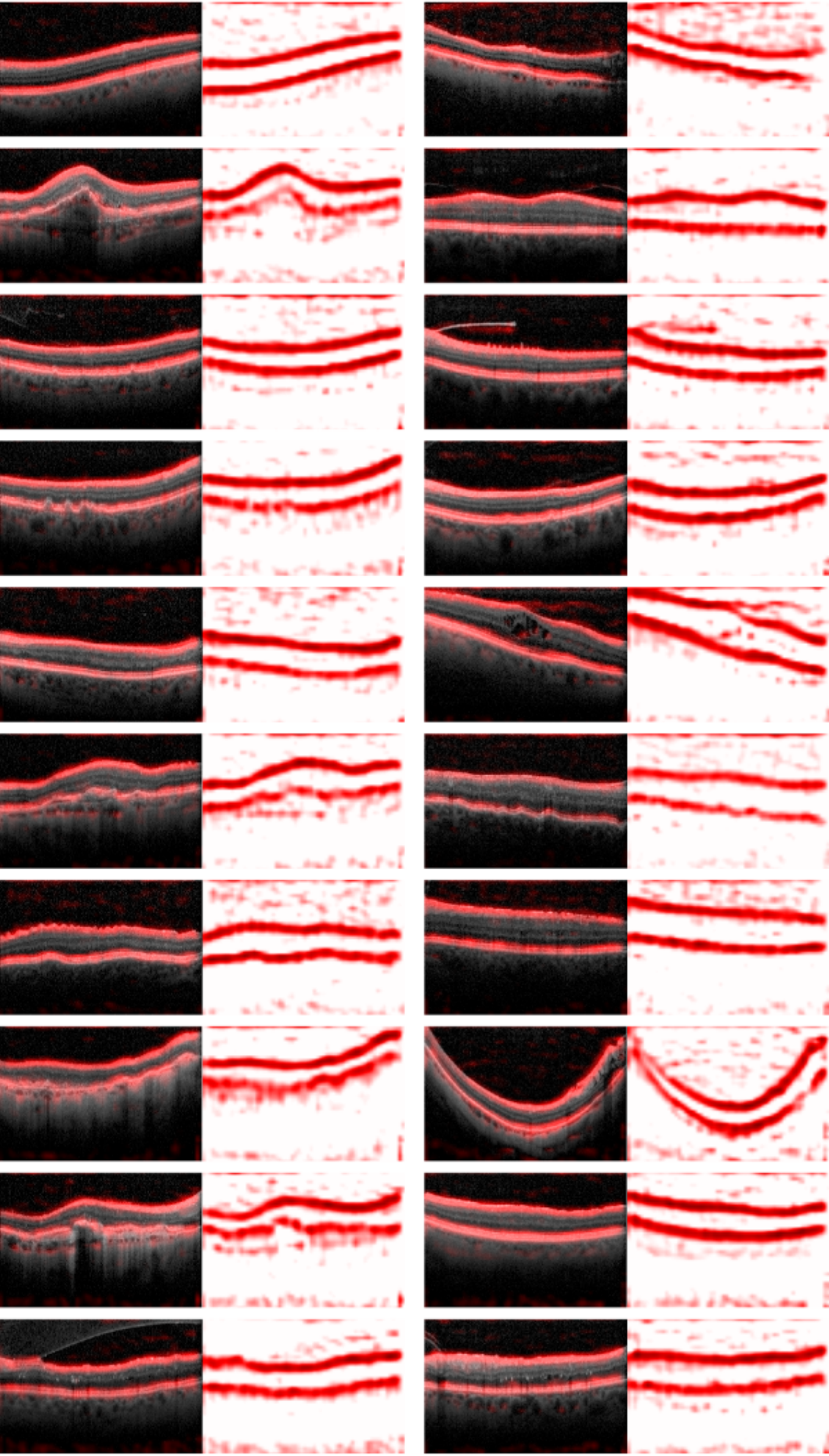}}
	\caption{Feature map 114 of layer 6 appears to have learned to track the Inner Limiting Membrane (ILM) and Retinal Pigment Epithelium (RPE).}
	\label{map114}
\end{figure}

\begin{figure}
\centering
	\makebox[0pt]{
	\includegraphics[scale=0.85]{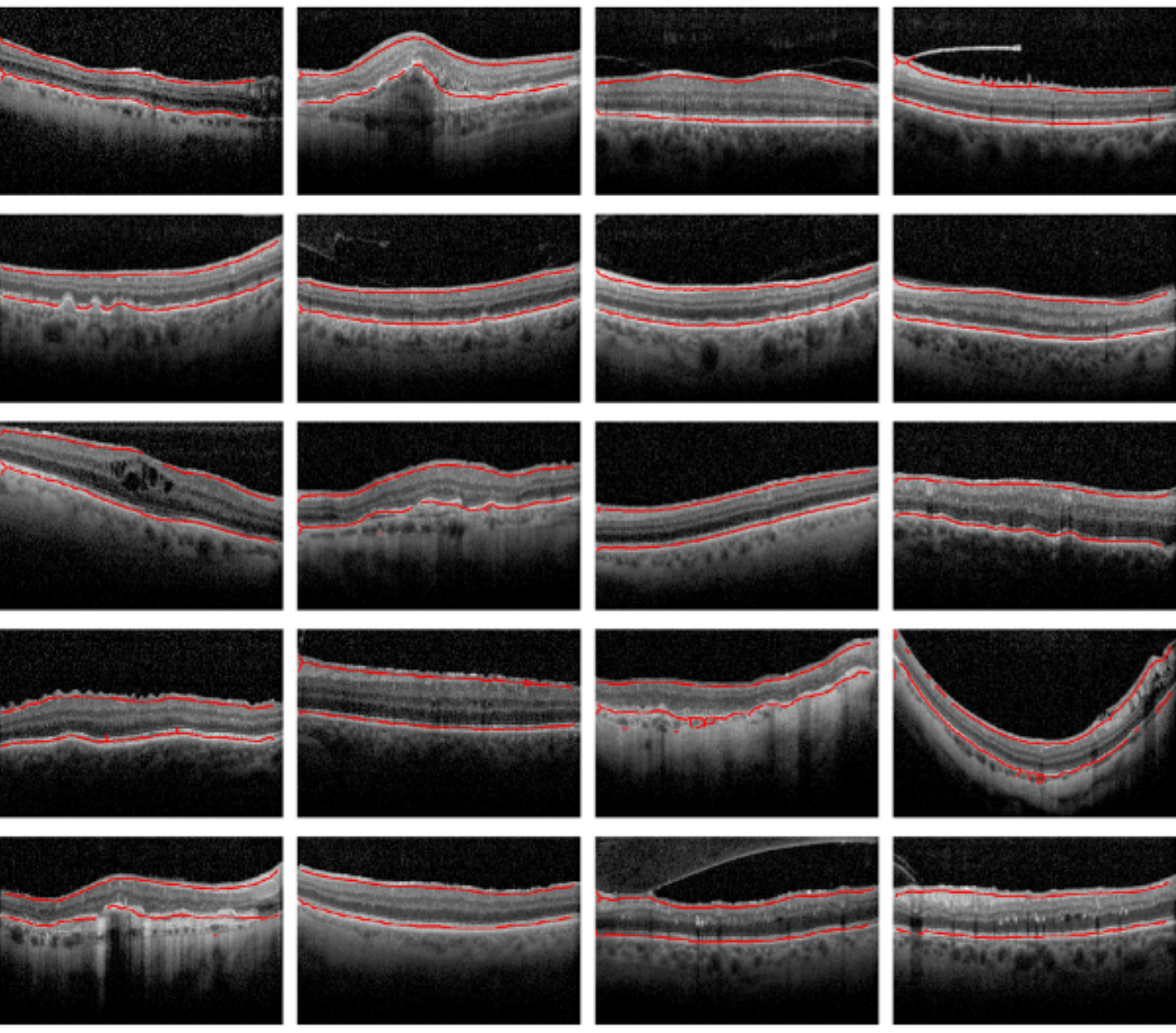}}
	\caption{Skeletonized version of the feature maps from Fig. \ref{map114} overlayed with their b-scans. The skeletons can be used to estimate retinal thickness, which can help in data selection or serve as an input feature for some other task. The feature maps were multiplied with the b-scan, thresholded by its mean value and median-filtered before skeletonizing.}
	\label{skeletons}
\end{figure}

\renewcommand{\arraystretch}{1.5}
\begin{table}
\begin{center}
\caption{Architecture of the HDcycleGAN generator. All layers use Instance Norm and ReLU activation. Skip connections link down-sampling and up-sampling layers using concatenation.}
\label{generator}
\begin{tabularx}{\textwidth}{l|Y|Y|Y}
\toprule
layer name	&	type	&  properties & output size\\
\midrule
initial convolution & convolution & kernel=7x7x16, stride=1 & 512x512x16 \\ \hline
down-sampling 1 & strided convolution & kernel=3x3x32, stride=2 & 256x256x32\\ \hline
down-sampling 2 & strided convolution & kernel=3x3x64, stride=2 & 128x128x64\\ \hline
down-sampling 3 & strided convolution & kernel=3x3x128, stride=2 & 64x64x128\\ \hline
residual block 1 & residual block & convolutions=3, kernel=3x3x128, stride=1 & 64x64x128\\ \hline
residual block 2 & residual block & convolutions=3, kernel=3x3x128, stride=1 & 64x64x128\\ \hline
residual block 3 & residual block & convolutions=3, kernel=3x3x128, stride=1 & 64x64x128\\ \hline
residual block 4 & residual block & convolutions=3, kernel=3x3x128, stride=1 & 64x64x128\\ \hline
residual block 5 & residual block & convolutions=3, kernel=3x3x128, stride=1 & 64x64x128\\ \hline
residual block 6 & residual block & convolutions=3, kernel=3x3x128, stride=1 & 64x64x128\\ \hline
up-sampling 1 & bilinear up-scaling + convolution & kernel=3x3x64, scale=2 & 128x128x64\\ \hline
up-sampling 2 & bilinear up-scaling + convolution & kernel=3x3x32, scale=2 & 256x256x32\\ \hline
up-sampling 3 & bilinear up-scaling + convolution & kernel=3x3x16, scale=2 & 512x512x16\\ \hline
final convolution & convolution & kernel=3x3x1, stride=1 & 512x512x1\\
\bottomrule
\end{tabularx}
\end{center}
\end{table}

\begin{table}
\begin{center}
\caption{Architecture of the HDcycleGAN discriminator. All layers use Instance Norm and ReLU activation, except the final layer which uses softmax and no norm.}
\label{discriminator}
\begin{tabularx}{\textwidth}{l|Y|Y|Y}
\toprule
layer name	&	type	&  properties & output size\\
\midrule
down-sampling 1 & strided convolution & kernel=3x3x16, stride=2 & 256x256x16\\ \hline
residual block 1 & residual block & convolutions=2, kernel=3x3x16, stride=1 & 256x256x16\\ \hline
down-sampling 2 & strided convolution & kernel=3x3x32, stride=2 & 128x128x32\\ \hline
residual block 2 & residual block & convolutions=2, kernel=3x3x32, stride=1 & 128x128x32\\ \hline
down-sampling 3 & strided convolution & kernel=3x3x64, stride=2 & 64x64x64\\ \hline
residual block 3 & residual block & convolutions=2, kernel=3x3x64, stride=1 & 64x64x64\\ \hline
down-sampling 4 & strided convolution & kernel=3x3x128, stride=2 & 32x32x128\\ \hline
residual block 4 & residual block & convolutions=2, kernel=3x3x128, stride=1 & 32x32x128\\ \hline
down-sampling 5 & strided convolution & kernel=3x3x256, stride=2 & 16x16x256\\ \hline
residual block 5 & residual block & convolutions=2, kernel=3x3x256, stride=1 & 16x16x256\\ \hline
down-sampling 6 & strided convolution & kernel=3x3x512, stride=2 & 8x8x512\\ \hline
residual block 6 & residual block & convolutions=2, kernel=3x3x512, stride=1 & 8x8x512\\ \hline
down-sampling 7 & strided convolution & kernel=3x3x1024, stride=2 & 4x4x1024\\ \hline
residual block 7 & residual block & convolutions=2, kernel=3x3x1024, stride=1 & 4x4x1024\\ \hline
average pooling & per channel averaging & & 1024 \\ \hline
logits & fully connected & & 3 \\
\bottomrule
\end{tabularx}
\end{center}
\end{table}
\end{document}